\documentclass[aip,floatfix]{revtex4-1}
\usepackage[dvips]{graphicx}
\usepackage{color}
\usepackage{amsmath,amssymb}
\usepackage[caption=false]{subfig}
\usepackage{epstopdf}

\begin{document}
\title{Efimov-DNA Phase diagram: three stranded DNA on a cubic lattice}
\author{Somendra M. Bhattacharjee}
\email{somendra.bhattacharjee@ashoka.edu.in}
\affiliation{Department of Physics, Ashoka University, Sonepat, 131029 India}
\author{Damien Paul Foster}
\email{damien.foster@coventry.ac.uk}
\affiliation{Centre for Computational Science and Mathematical Modelling, Coventry University, Coventry, UK CV1 5FB}

\begin{abstract}

We define a generalised model for three-stranded DNA  consisting of two chains of one type and a third chain of a different type. 
The DNA strands are modelled by random walks on the  three-dimensional cubic lattice with different interactions between two chains of the same type and two chains of different types. This model may be thought of as a classical analogue of the quantum three-body problem.
  In the quantum situation it is known that three identical quantum particles will form
  a triplet with an infinite tower of bound states at the point where
  any pair of particles would have zero binding energy.  The phase diagram is mapped out, and the different phase transitions examined using finite-size scaling. We look particularly at the scaling of the DNA model at the equivalent Efimov point for chains up to 10000 steps in length. We find clear evidence of several bound states in the finite-size scaling. We compare these states with the expected Efimov behaviour. 
  
 \end{abstract}

\maketitle

\section{Introduction}
One of the strange results in quantum mechanics is the formation of an
infinite number of bound states in a three-particle system when any
two would have given a zero-energy bound state.  This result goes by
the name of Efi\-mov
effect\cite{efimov1,efimov2,Braaten,kraem,zacca,pires}.  It has
recently been argued that the classical analogue of the Efi\-mov
effect is the formation of triple-stranded DNA at the melting point
of duplex DNA\cite{maji1}.  In this paper we introduce a generalised
three-stranded DNA model and examine its phase diagram using the
flatPERM Monte Carlo method.  Our model consists of a simple extension
of the usual Gaussian-Chain model of DNA\cite{causo}.  In the standard
DNA model, the configurations of two identical random walk chains of
given length, joined at a common origin, are considered where the only
energy comes from base-pairings (common-visited sites) occurring 
at the same contour length from the common origin
along both chains\cite{watcri}.

\begin{figure}
\begin{center}
\includegraphics[height=10cm]{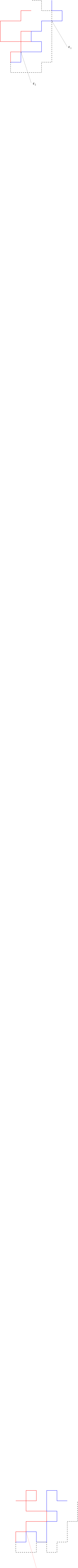}
\end{center}
\caption{The type A chain is shown as a dashed line, whilst the type B chain are shown as solid lines. The only two interactions occur at equal contour distances from the common origin. All other contacts do not give rise to interactions.}\label{model}
\end{figure}

In our model we label the two chains from the standard model as
 type {\bf B}, and introduce a third chain, which we  label 
type {\bf A}. We  denote the interaction strength by dimensionless
variables (i.e. absorbing the factor $k_BT$ into the interaction
strength,{{ where $k_B$ is the Boltzmann constant and $T$ is the
temperature}}) 
$\varepsilon_1$ for a base-pairing interaction between the
 type {\bf A} chain and either of the   type {\bf B} chains, and
$\varepsilon_2$ for base-pairing interaction between the two chains of
type {\bf B}.  These are the only interactions included; there are no
three-chain interactions other than those generated between the chains
in pairs. The model is shown for strands of length of 14 on the square lattice in Fig~\ref{model}.

{{Our purpose in this paper is to 
examine the full phase diagram }}for the model defined through the 
partition function
\begin{equation}
Z_N(\varepsilon_1,\varepsilon_2)=\sum_{\Omega_{3,N}} g_{3,N}(n_1,n_2) \exp(n_1\varepsilon_1+n_2\varepsilon_2),
\end{equation}
where we have denoted by $\Omega_{3,N}$ the set of configurations of
three random walks of length $N$. Here, $g_{3,N}(n_1,n_2)$ is the number of
configurations with exactly $n_1$ interactions between the
type {\bf A} chain with either of the type {\bf B} chains
($n_1\in[0,2N]$) and exactly $n_2$ interactions between the chains of
type {\bf B}.  The attractive contact energies are taken as
$\varepsilon_1$ and $\varepsilon_2$ for  {\bf A}-{\bf B} and {\bf
  B}-{\bf B} pairs.  Note that the temperature has been absorbed in
the definition of these contact energies, so that $\varepsilon_{1,2}$
are dimensionless. In other words, the contact energies are given by
$k_BT \varepsilon_{1,2}.$  {{We also examine the scaling
of the free-energy at the three-stranded DNA equivalent to the Efimov
point.}} 

The Efimov point is located where the inter-chain interactions are the
same, and at a value where any two chains are at the two-chain binding
transition.  



\section{Connection between DNA model and the Efi\-mov effect}
The formal connection between DNA melting and the quantum problem can be
established as follows\cite{maji1}.  Take three gaussian polymers with native
base-pairing interaction, i.e., two monomers on two chains interact if
and only if they have the same contour length index measured from a
predetermined end.  The Hamiltonian is
\begin{eqnarray}
  \label{eq:1}
  {\sf H}=\frac{H}{k_BT}
  &=&\sum_{i=1,2,3} \frac{1}{2} \int_0^N \left(\frac{\partial
      {\bf r}_i(s)}{\partial s}\right )^2 ds +\nonumber\\
&&\qquad\sum_{i<j}  \int_0^N V({\bf r}_i(s) -
{\bf r}_j(s)) ds,
\end{eqnarray}
where ${\bf r}_i(s)$ is the position coordinate of a monomer (or base)
at contour length $s$.  The first term represents the elastic energy or the
connectivity of the chain as a polymer, while the second is the
interaction between two monomers at the same contour length $s$
(native base pairing of DNA).  Like a Hydrogen bond, the range of
interaction of $V$ is taken to be small.  The partition function is
then given by
\begin{equation}
  \label{eq:2}
  Z=\int {\cal DR} \exp(-{\sf H}),
\end{equation}
where the integral represents the sum over all configurations as a
path integral.

If we now do an imaginary transformation $s=it$, then the
partition function changes to
\begin{subequations}
\begin{eqnarray}
  \label{eq:3}
  G&=& \int {\cal DR}\, \exp(iS),\\
{\rm \, with, \,}\   S&=&\int dt L,\\
  {\rm and,}\,\,\ L&=&\sum_{i=1,2,3} \frac{1}{2} 
      \left(\frac{\partial{\bf r}_i(t)}{\partial t}\right )^2 -
        \sum_{i<j} V_{ij},\label{eq:9}
\end{eqnarray}
\end{subequations}
as if $L$ represents the Lagrangian of three particles with pairwise
interaction $V_{ij}=V({\bf r}_i(t) -{\bf r}_j(t))$, with $t$ as real
time.  In this path-integral representation, $G$ now describes the
quantum propagator of three particles with short-range, pairwise
interactions.  The key point in this exact transformation is the
native base pairing of DNA (monomers with the same index) that got
translated into the same time interaction in the quantum picture. In
the $N\to\infty$ limit, the groundstate energy of the quantum problem
maps onto the free energy of the DNA problem.

Double-stranded DNA undergoes a melting transition, as temperature is
increased, or as the strength of the potential in Eq. (\ref{eq:1}) is
changed.  The melting point corresponds to the critical strength of
the potential in the quantum problem above, in  which a bound state occurs
in a short-range potential in three dimensions.  The bound state
energy is related to the width $a$ of the wave function,  with $E\sim
\hbar^2/2ma^2$, so that $E\to 0$ implies $a\to \infty$ \cite{scatt}.
 
At the critical value of zero energy bound state, Efimov argued that
three particles will produce a long-range effective interaction
$-1/r^2$ which leads to a tower of bound states with energies
\begin{equation}\label{eq:10}
E_n=\gamma^n E_0, E_0<0, ~ {\rm where} ~ \gamma\approx \frac{1}{(22.7)^2}.   
\end{equation}
This is the Efimov effect.

The quantum fluctuations arise from the paths in the classically
forbidden regions which are outside the potential well.  In the DNA
picture, these are the regions on the chains where the hydrogen bonds
are broken by thermal fluctuations.  A portion of the duplex with
broken hydrogen bonds will 
be called a bubble.  The bubbles are characterized by two lengths,
$\xi_{\parallel}$, the fluctuation in the number of bonds broken, and
$\xi$ the corresponding length scale for the spatial size, with the
relation
\begin{equation}
  \label{eq:6}
\xi \sim \xi_{\parallel}^{\nu},
\end{equation}
where $\nu$ is the polymer size exponent.  For Gaussian polymers
(random walks) $\nu=1/2$.

The melting transition of the DNA at temperature $T=T_c$, where the
hydrogen bonds of the duplex DNA are cooperatively broken, is
described by the free energy per unit length
\begin{equation}
  \label{eq:4}
f\sim  \frac{k_BT_c}{\xi_{\parallel}}.
\end{equation}
For a continuous transition,
as one finds from exact solutions or from the Poland-Scheraga
arguments, we may take $f\sim (|T-T_c|/T_c)^{2-\alpha}$ at least for
$T<T_c$, so that $\xi_{\parallel}\sim (|T-T_c|/T_c)^{\alpha-2}$.  This
transition, like many other critical points, shows continuous scale
invariance in the sense that under a scale transformation $r\to b r$,
the free energy scales as $f\to b^{-2} f$ for any $b$.

Let us now consider two strands of DNA --- let us call them A, B ---
separated by a distance $R$ much larger than the hydrogen bond
distance so that they do not form any doublet.  Now we add a third
chain C that can pair with both A and B with the same bond energy. If
the temperature is close to the melting point of any pair, the large
bubbles will allow C to make contacts with both A and B, resulting in
an attraction between the latter two.  This fluctuation induced
interaction is given by an $R$-dependent free energy, modifying Eq.
(\ref{eq:4}) to
\begin{equation}
  \label{eq:5}
  \Delta f\sim - \frac{k_BT_c}{\xi_{\parallel}} {\cal F}(R/\xi),
\end{equation}
where the scaling function ${\cal F}(x)$ should be such that
Eq.(\ref{eq:5}) makes sense for
$\xi_{\parallel},\xi\to\infty$.  By using Eq. (\ref{eq:6}), we then
require ${\cal F}(x)\sim x^{-1/\nu}$ so as to cancel
$\xi_{\parallel}$. At the critical point for C, we then get
\begin{equation}
  \label{eq:7}
  \Delta f(R)\sim -\frac{1}{R^2}, 
\end{equation}
where the gaussian chain value $\nu=1/2$ has been used.

The above long-ranged inverse-square interaction is at the heart of
the Efimov effect, but it is obtained here via the DNA mapping.  For
DNA, this interaction would lead to a bound phase of three strands at
the melting point of the duplex DNA.  Consequently, the three chain
complex will melt at a temperature higher than $T_c$.

There are two aspects of the Efimov effect.  One is the formation of a
three-particle bound state for potentials where two would not have
formed a bound state.  The second one, more subtle, is the formation
of the Efimov tower precisely at the critical potential of zero energy
bound-state for a pair,  corresponding to the breaking of the
continuous scale invariance of the critical point to a discrete scale
invariance\cite{hammer,pal}.

\section{Phase Diagram}

In this section we present numerical results for the phase diagram,
where we examine the different transition lines and phases. In order
to determine the phase diagram, we have used the flatPERM
method\cite{perm} to stochastically enumerate (or partially enumerate)
the coefficients of the relevant partition functions.

For the full model, we can define the partition function $Z_N$ through:
\begin{equation}
Z_N(\varepsilon_1,\varepsilon_2)=\sum_{\Omega_{3,N}} g_{3,N}(n_1,n_2) \exp(n_1\varepsilon_1+n_2\varepsilon_2),
\end{equation}
where we have denoted by $\Omega_{3,N}$ the set of configurations of
three random walks of length $N$. $g_{3,N}(n_1,n_2)$ is the number of
configurations with exactly $n_1$ interactions between the chain of
type {\bf A} with either of the type {\bf B} chains ($n_1\in[0,2N]$)
and exactly $n_2$ interactions between the type {\bf B} chains.

\begin{figure}[t]
\vspace{0.3cm}
\begin{center}
\includegraphics[width=\linewidth]{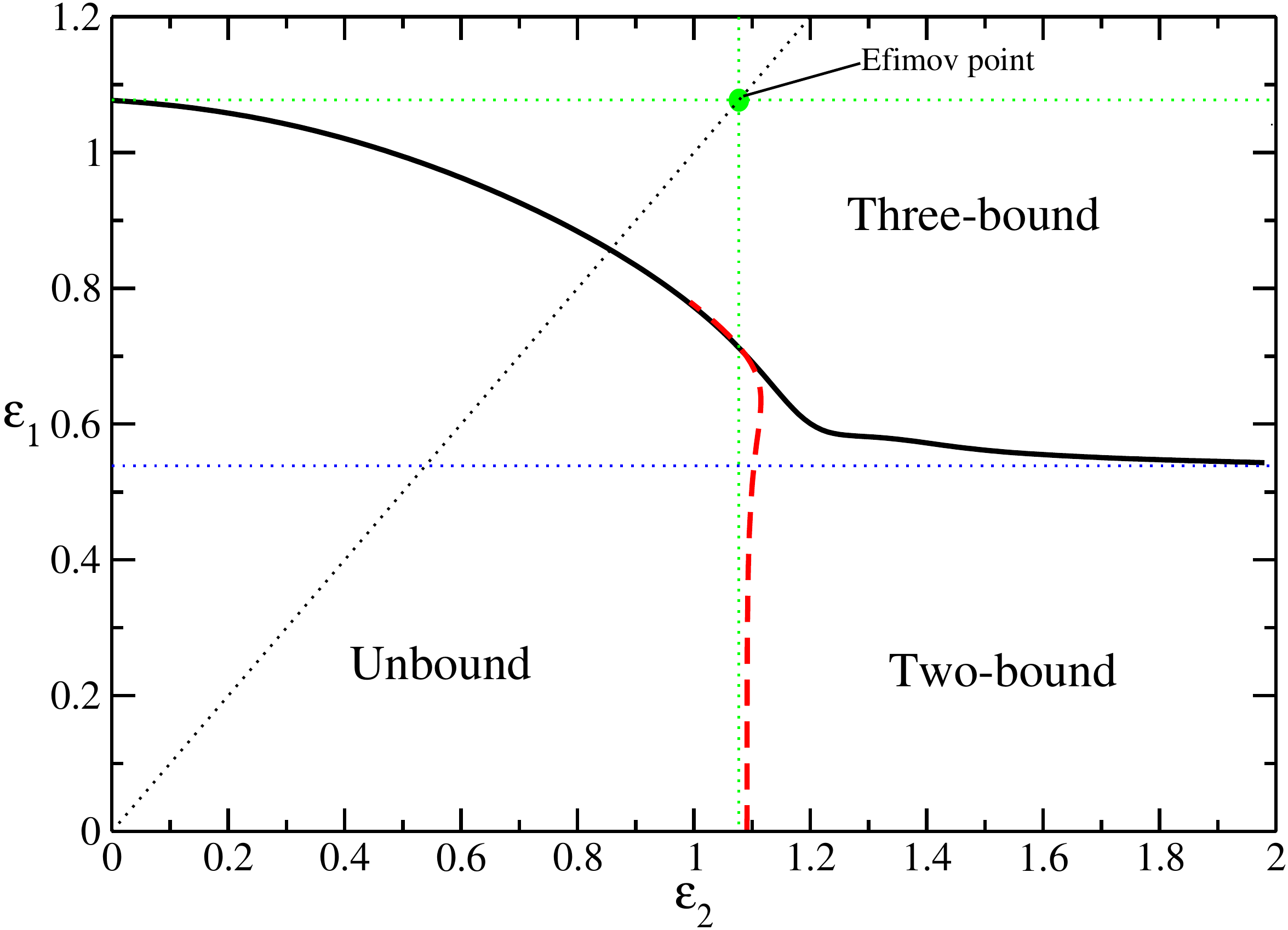}
\end{center}
\caption{Phase diagram in the $\epsilon_1$-$\epsilon_2$ plane.  The
  dotted horizontal and vertical lines at $\epsilon_c^{(2)}=1.07726...$ are
  the two-chain melting lines.  The horizontal line at 0.5357... is
  the expected transition line for peeling of A from tightly bound BB
  pair. The solid line represents the transition to the three chain bound
  state, see text for details on the nature of the transition. The
  $\epsilon_1=\epsilon_2$ line meets the two-chain melting lines at the
  Efimov point (green disk). 
  There is a region where three chains are bound though no two should
  have been bound.  This region is the Efimov-DNA
  region.}\label{pdfig}
\end{figure}

Having good estimates of $g_{3,N}(n_1,n_2)$ then allows densities and
fluctuations in energy to be calculated directly. Suppose we have a
quantity $X(n_1,n_2)$, which we also calculate during the flatPERM
calculation, then the average is calculated:
\begin{equation}
\langle X \rangle =   \frac{\sum_{\Omega_{3,N}} X(n_1,n_2) g_{3,N}(n_1,n_2) e^{n_1\varepsilon_1+n_2\varepsilon_2}}{Z_N(\varepsilon_1,\varepsilon_2)}.
\end{equation}
In particular, we can calculate the average number of contacts $\langle n_{i}\rangle$ and the corresponding fluctuations \mbox{$\Delta_i=\frac{1}{N}\left(\langle n_i^2\rangle-\langle n_i\rangle^2\right)$}, with $i=1, 2$.

The average number of contacts is expected to scale as\cite{causo} 
\begin{equation}
\langle n_i \rangle \sim N^{\phi_i},
\end{equation}
where $\phi_i=0$ in the unbound phase and $\phi_i=1$ in the bound
phase, and taking a potentially non-trivial value at the transition.
This behaviour enables the setting up of a phenomenological
renormalisation group method\cite{night1976} using the function
\begin{equation}\label{varphis}
\varphi_{i,N,N^\prime} = \frac{\log(\langle n_i \rangle_N/ \langle n_i \rangle_{N^\prime})}{\log(N/N^\prime)}.
\end{equation}
Estimates for the critical values of $\varepsilon_1$ may be calculated
looking for crossings of the $\varphi_{i,N,N^\prime}$ keeping
$\varepsilon_2$ {{fixed}} (and the other way round). These
crossings give estimates of $\phi_i$ at the transition. Logically, one
uses $\varphi_{1,N,N^\prime}$ to calculate the critical values
$\varepsilon^\star_1(\varepsilon_2)$ (and vice-versa). The solid black
line in the phase diagram in Fig~\ref{pdfig} is calculated from
$\varphi_{1,N,N/2}=\varphi_{1,N/2,N/4}$ and the red dashed line using
$\varphi_{2,N,N/2}=\varphi_{2,N/2,N/4}$ with $N=200$.

The phase diagram consists of three distinct phase transition lines
that join at a multi-critical point and separate out three phases:
unbound, two-bound and three-bound, corresponding to the number of
chains involved in the bound states.

\subsection{Unbound/Two-Bound Phase boundary} 

When $\epsilon_1=0$ the type {\bf A} chain does not interact at all
with the two type {\bf B} chains, and the transition as
$\varepsilon_2$ is decreased is the standard two chain DNA melting
transition at $\varepsilon_2=\varepsilon_c^{(2)}$, where
$\varepsilon_c^{(2)}=1.07726\cdots$ is the 2-chain binding
transition\cite{causo}.

As $\varepsilon_1$ is increased, we can view the situation of the
2-chain complex adsorbing to the type {\bf A} chain. Whilst the number
of contacts is small ($\phi_1=0$), the chain type {\bf B} will not
affect the two-chain binding transition, and we would expect
$\varepsilon_{2,c}(\varepsilon_1)=\varepsilon_c^{(2)}$ to remain
constant until the third chain binds at the multicritical point. In
Fig~\ref{pdfig} the discrepancy between the estimated line (dashed)
and expected transition is due to finite size effects. On the
transition line we expect $\phi_1=0$ and $\phi_2=1/2$, which is
consistent with the results found for chains up to $N=200$.

\begin{figure}
\includegraphics[width=\linewidth]{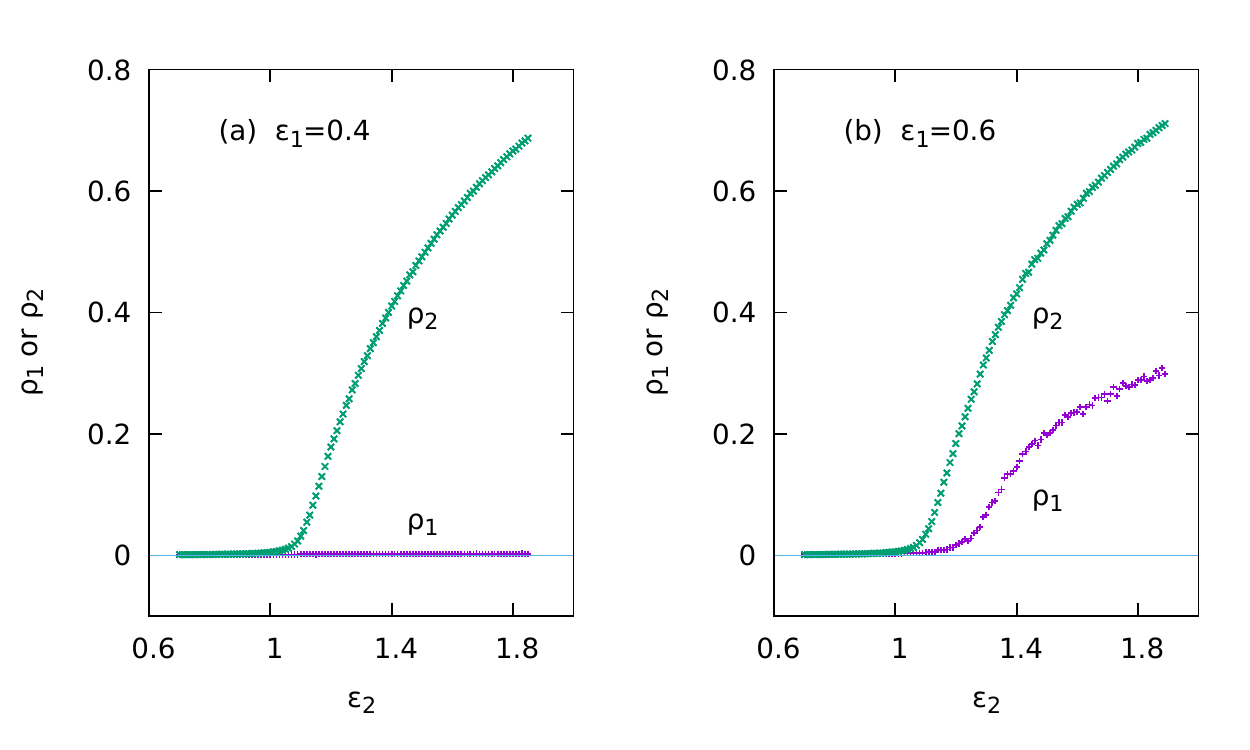}
\caption{Plots of $\rho_1,\rho_2$ vs $\epsilon_2$ for (a)
  $\epsilon_1=0.4$, and (b) $\epsilon_1=0.6$ for chains of length
  2000.  In (a) we see the Melting of BB at $\epsilon_c^{(2)}$ with
  decreasing $\epsilon_2$, while A remains unbound.  In (b) we see
  different melting points for BB (at $\epsilon_c^{(2)}$), and A-BB triplet.
}\label{fig:bbmelt}
\end{figure}

\subsection{Two-bound/Three-bound boundary}

As $\varepsilon_2\to\infty$, the two type {\bf B} chains become
tightly bound and behave as one Gaussian chain.  Each contact with the
type {\bf A} chain is a double contact, and we would thus expect a
binding transition when $\varepsilon_1=\varepsilon_c^{(2)}/2$. As
$\varepsilon_2$ is lowered, whilst the two type {\bf B} chains remain
bound, they will start containing bubbles.  Now, when the type {\bf A}
chain comes into contact with the bound duplex, the number of contacts
will sometimes be with one chain and sometimes with 2, making it
harder to bind.  This will have the effect of elevating the critical
temperature, or making it harder to bind, such that
$\varepsilon_{1,c}(\varepsilon_2)>\varepsilon_c^{(2)}/2$.  As the
$\epsilon_2\to\epsilon_c^{{(2)}}$, the phase transition line merges with
the 2-chain binding transition at the multi-critical point. Along this
line we expect $\phi_1=1/2$, as this is a standard type binding
between two random walks (at least for large $\varepsilon_2$, and we
see no evidence of a change in behaviour before the multi-critical
point) and $\phi_2=1$, since the two type {\bf B} chains are bound.
This is also borne out by the numerical results. In
Figure~\ref{fig:bbmelt} we show the plots of $\rho_1$ and $\rho_2$ as
a function of $\epsilon_2$ for two values of $\varepsilon_1$. When
$\varepsilon_1=0.4$ (Fig.~\ref{fig:bbmelt}~(a)) we see that the
density $\rho_1$ remains zero, whilst $\rho_2$ becomes non-zero as the
phase boundary is crossed. When $\varepsilon_1=0.6$
(Fig.~\ref{fig:bbmelt}~(a)) $\rho_1$ becomes non-zero later than
$\rho_2$, as we cross successively the unbound/two-bound phase
boundary and the two-bound/three-bound phase boundary.

\subsection{Unbound/Three-bound boundary}

 We first consider the case where $\varepsilon_2=0$.
 When $\varepsilon_2=0$, the two type {\bf B} chains do not see each other, they only see the  type {\bf A} chain. At the critical interaction $\varepsilon_1=\varepsilon_c^{(2)}$ each type {\bf B} chain binds with the chain of type {\bf A} independently. This transition can be seen as two independent events. The chain of type {\bf A} binds the other two into a triplex-bound state. This bound state ensures that the two chains of type {\bf B} remain close to each other. The number of interactions $n_1$ is the sum of the number of interactions with each of the chains of type {\bf B}, and these contacts will be decorrelated between the two chains. It is clear then that $\langle n_1\rangle\sim N^{1/2}$, giving $\phi_1=1/2$.

\begin{figure}
 \subfloat[$\varepsilon_2=0$]{\includegraphics[width=0.5\linewidth]{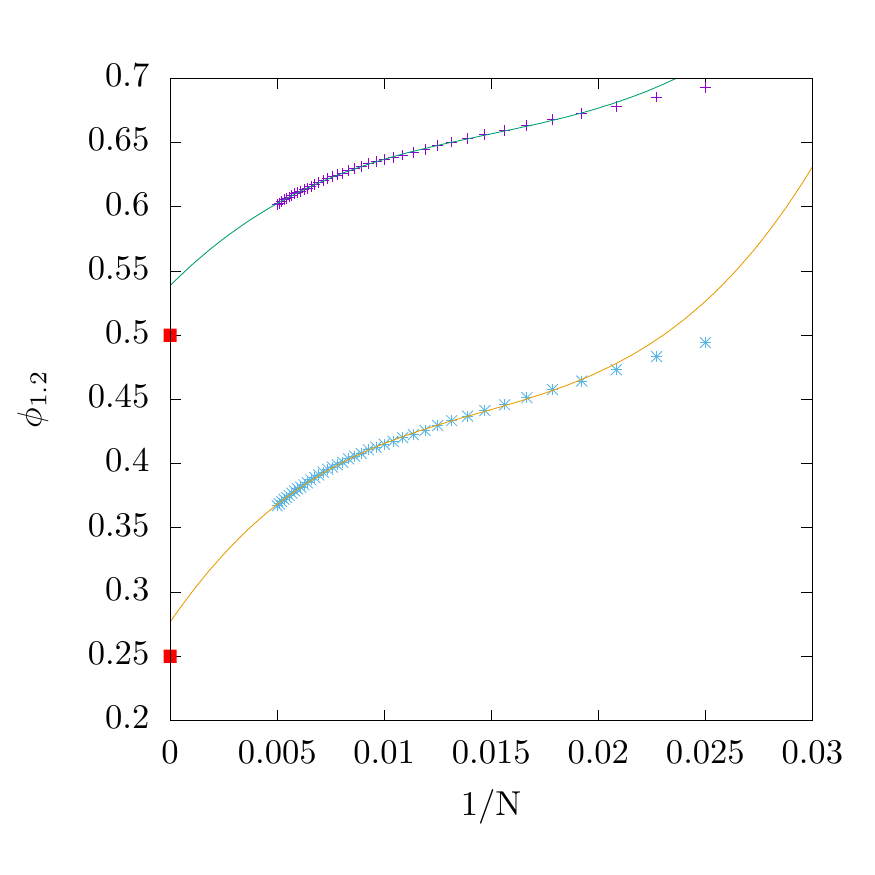}}
 \subfloat[$\varepsilon_2=0.7$ and $\varepsilon_1=\varepsilon_2$]{\includegraphics[width=0.5\linewidth]{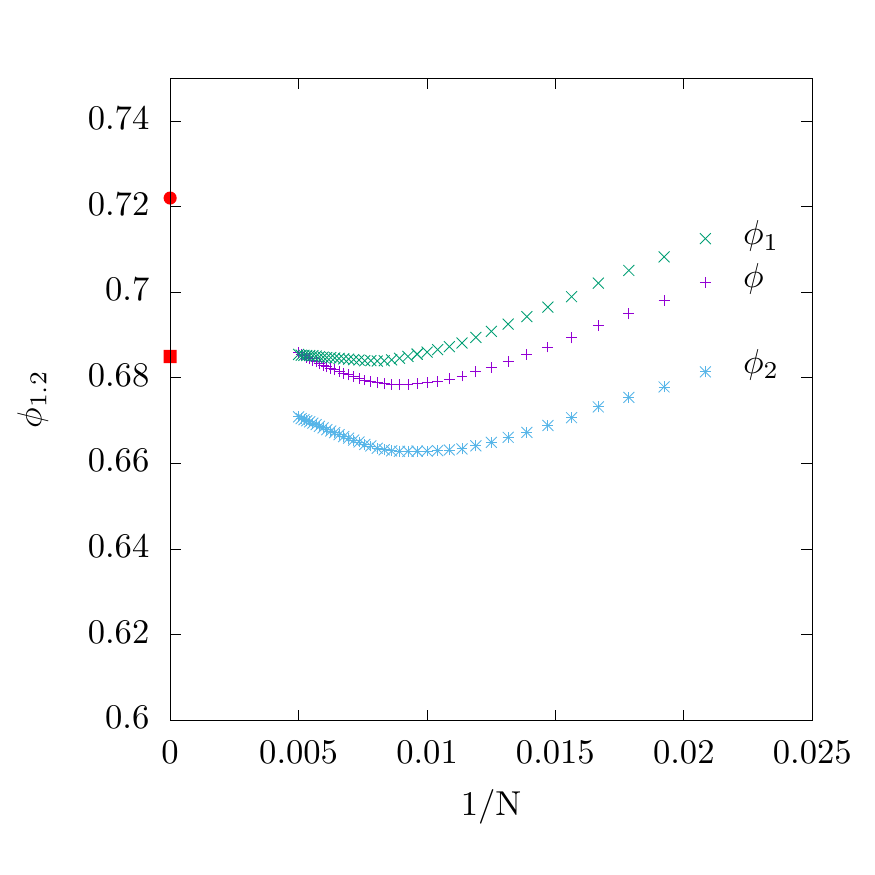}}
 \caption{In figure (a) we show plots of $\phi_1$ and $\phi_2$ for $N\leq 200$ plotted against $1/N$.  Figure (b) shows $\phi_1$ and $\phi_2$ for $\varepsilon_2=0.7$ and $\phi=\phi_1=\phi_2$ for $\varepsilon_1=\varepsilon_2$.}\label{phis}
 \end{figure}

Estimates of $\phi_1$ and $\phi_2$ are shown in Figure~\ref{phis}, extrapolated by fitting to a quadratic function. It can be seen that $\phi_1$ could reasonably extrapolate to $1/2$, and $\phi_2$ to a non-zero value, possibly $1/4$, which is consistent with there being a bound triplet state, where the two type {\bf B} chains are bound through the intermediary of type {\bf A} chain.
 
 We looked at the free energy for the triplet state at this point, and found it to be the same form as the free energy for the two-chain DNA model at its melting point (shown in Fig.~\ref{fig:2crit}) but twice as large.

 \begin{figure}
\includegraphics[width=\linewidth]{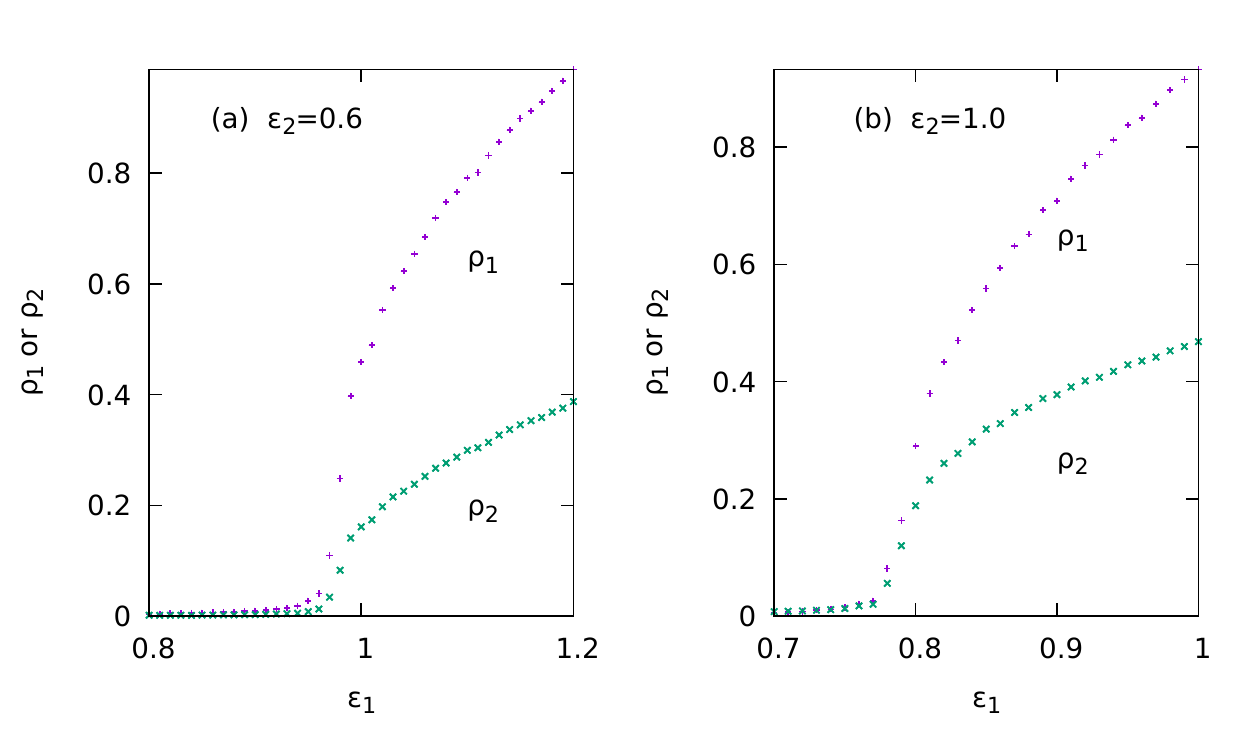}
\caption{Efimov-DNA. Plots of $\rho_1,\rho_2$ vs $\epsilon_1$ for (a)
  $\epsilon_2=0.6$, and (b) $\epsilon_2=1.0$ for chains of length
  2000.  We see both BB and AB pairings at the same
  $\epsilon_1<\epsilon_c^{(2)}$ for $\epsilon_2=0.6$ (a) and
  $\epsilon_2=1.0$ (b).  The transition is between the unbound and the
  ABB triplet phases.  The transition takes place in the region where
  any pair would have been in the unbound phase.  }\label{fig:efimelt}
\end{figure}

This is interesting, since it is clear that when $\varepsilon_1=\varepsilon_2$ the three chains are equivalent, and $\phi_1=\phi_2$, which indicates that the point $\varepsilon_1=\varepsilon_c^{(2)}, \varepsilon_2=0$ is different in nature from the rest of the line. This is understandable, because the type {\bf B} chains are already bound by the type {\bf A} chain, and so a small change in $\varepsilon$ could reasonably make a big change. In Fig.~\ref{fig:efimelt} we look at the densities of the interactions $\rho_1=n_1/N$ and $\rho_2=n_2/N$ as a function of $\varepsilon_1$ for $\varepsilon_2=0.6$ and $\varepsilon_2=1$. In both cases we can see that the two densities become non-zero at the same time, indicating clearly that the phase above the transition is a triplet phase.

Fig~\ref{phis}~(b) shows $\phi_1$ and $\phi_2$ for $\varepsilon_2=0.7$. The two $\phi$ seem as if they may reasonably give the same limiting value (around 0.68), which is bigger than $1/2$. We compare to the plot with $\varepsilon_1=\varepsilon_2$ (so $\phi_1=\phi_2$ by construction). Here we seem to have a different limit, leading to the possibility (unverified) that $\phi_{1,2}$ may vary along the unbound/3-bound phase boundary. 

There is another possibility, which is that the line is weakly first order, which is the prediction of studies on hierarchical lattices\cite{maji1}. It is difficult for the length of chains considered to tell the difference between a smooth variation of the density, or a small jump which might develop only for very long chains.

As $\varepsilon_2$ is increased, the  type {\bf B} chains will tend to bind more, which has the effect of making it easier for the type {\bf A} chain to bind, which lowers the value of $\varepsilon_1$ required to maintain the triple-bound state. Along the whole of this phase transition line, the two type {\bf B} chains are nevertheless held together by the action of the the type {\bf A} chain. This stops when $\varepsilon_2=\varepsilon_c^{(2)}$, and the {\bf B}-chains can bind in their own right. This occurs at the multi-critical point. 
 
 \subsection{The multi-critical point}
 
 \begin{figure}
\subfloat[]{\includegraphics[width=0.5\linewidth]{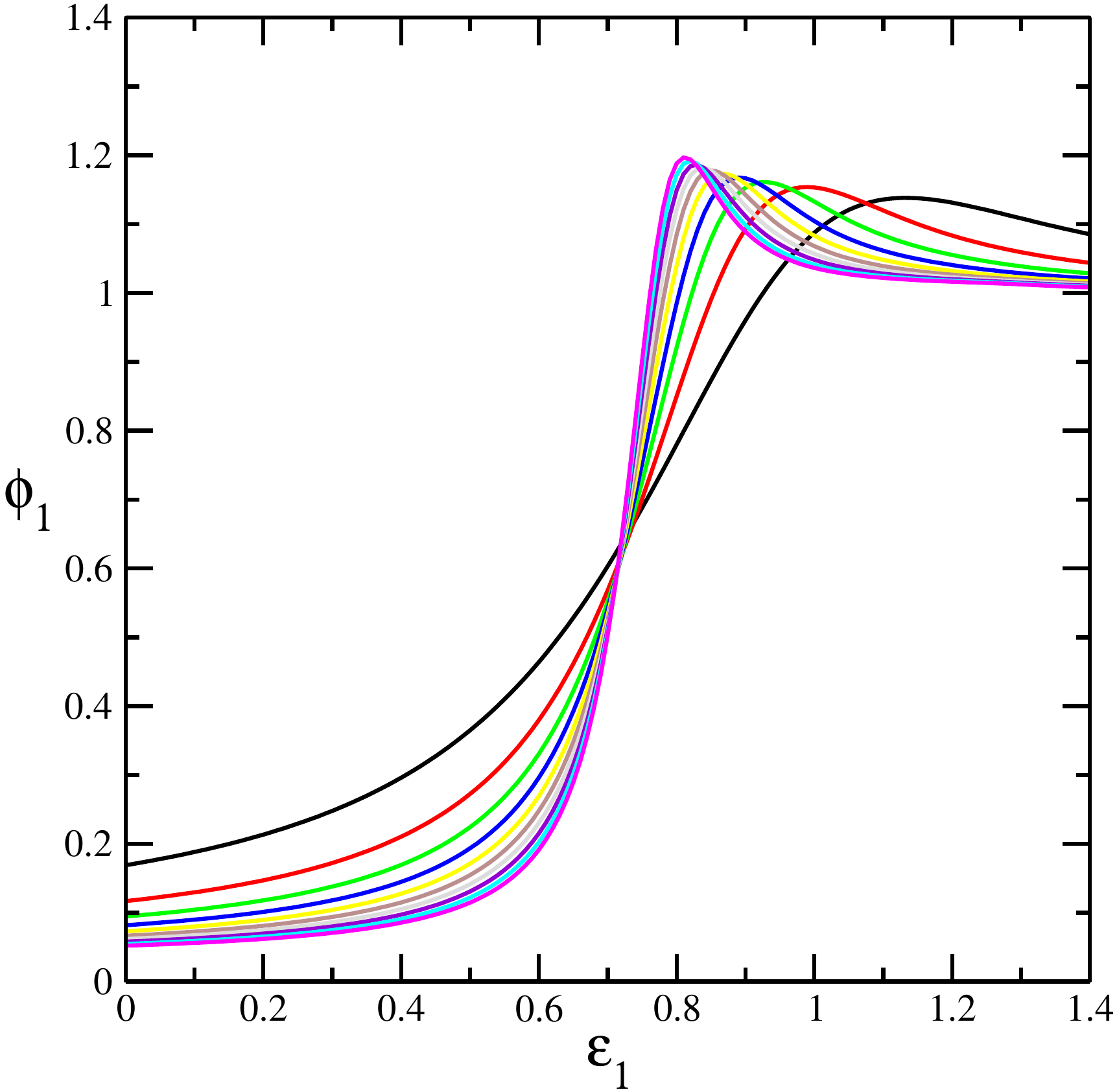}}
\subfloat[]{\includegraphics[width=0.5\linewidth]{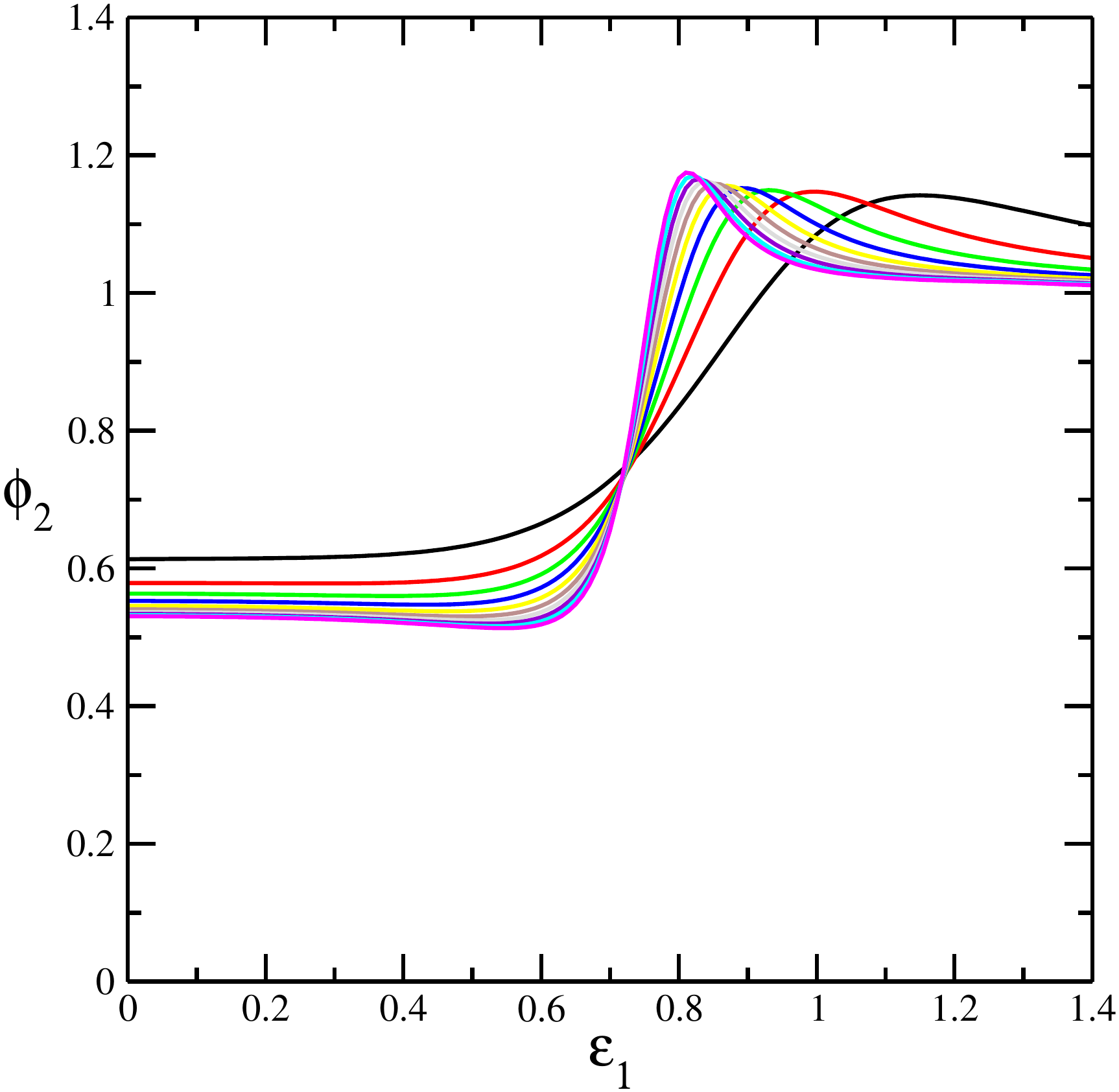}}\\
\subfloat[]{\includegraphics[width=0.5\linewidth]{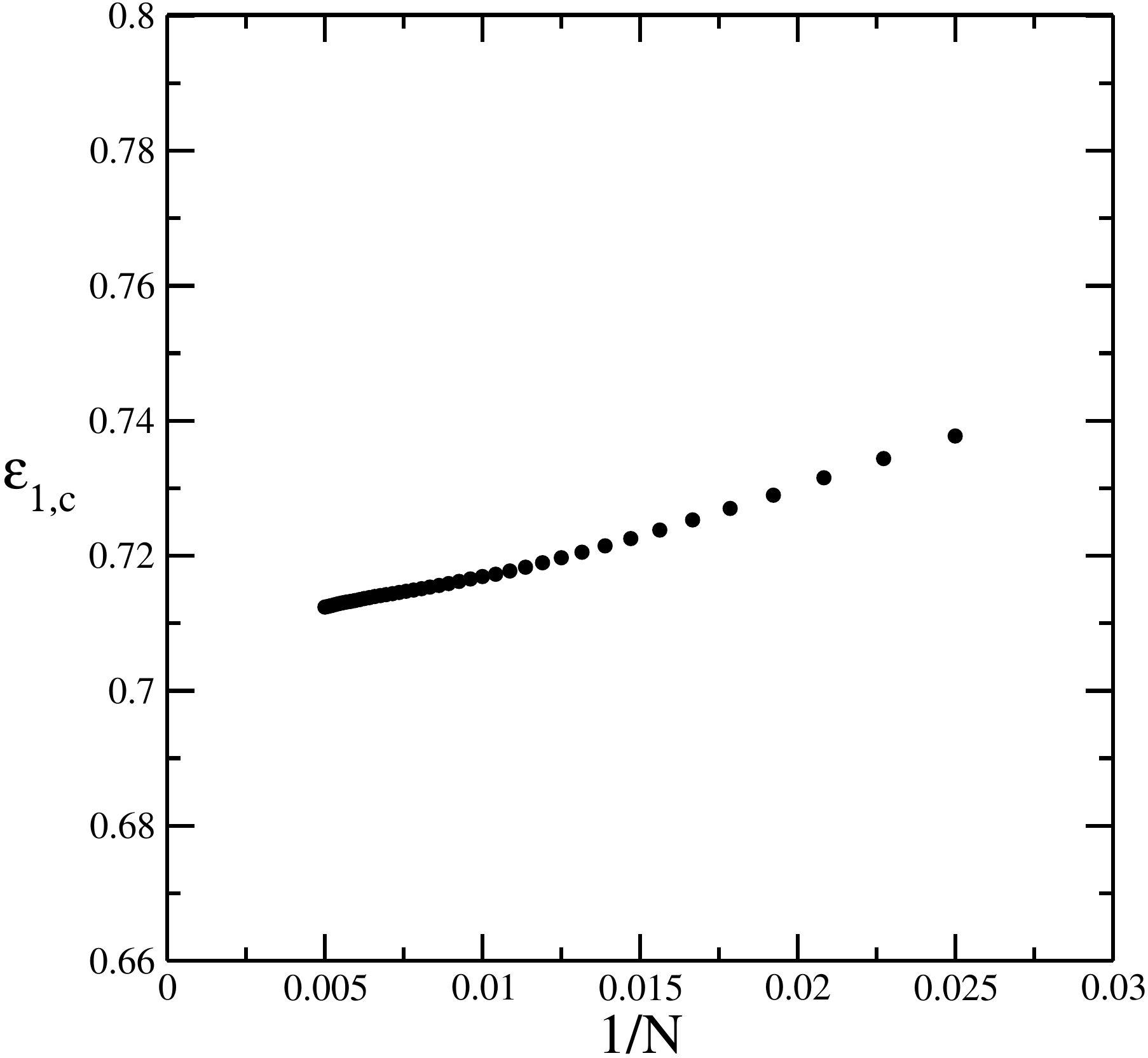}}
\subfloat[]{\includegraphics[width=0.5\linewidth]{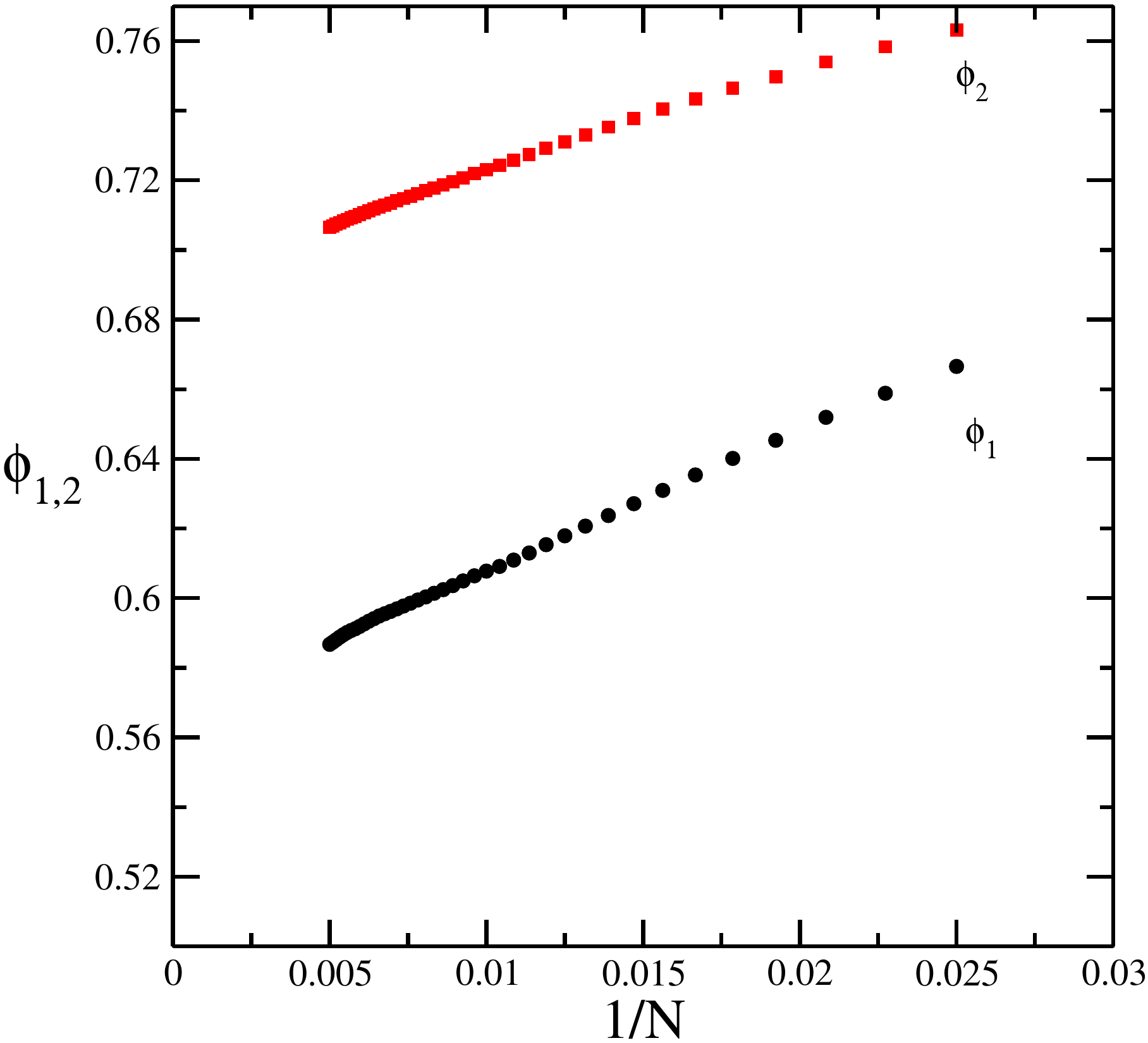}}
\caption{Figures (a) and (b) show plots of $\varphi_{1/2}$ vs $\varepsilon_1$. The crossings give finite-size estimates of the critical coupling. Figure (c) shows these estimates plotted against $1/N$ and Figure (d) shows estimates of the $\phi$ values. }\label{mpfind} 
\end{figure}

 The multi-critical point location can be identified by looking at the crossings for $\varphi_i$ defined in Eq.~\ref{varphis} along the line $\varepsilon_2=\varepsilon_c^{(2)}=1.07726\cdots$. The value of $\phi_1=0$ is expected along this line until the multicritical point, where it will be expected to take on a new value, indicating the adsorption of the type {\bf A} chain to the  type {\bf B} chains. Likewise, the value $\phi_2=1/2$ is expected along this line, but may or may not take on a new value at the multicritical point.  In Fig.~\ref{mpfind} we show plots of $\varphi_{1,2}$ against $\varepsilon_1$ showing crossings at the estimated location of the multicritical point. We show $\varphi_{1/2,N,N/2}$ for chains chains of lengths from $N=20$ to $N=200$ in steps of 20. In the same figure we show the variation of the estimate for the value of $\varepsilon_1$ at the multi-critical point, which we estimate to be at $\varepsilon_1=0.71(5), \varepsilon_2=1.07726...$.

\section{Finite-size scaling}

The distinctive feature of the Efimov effect is the occurrence of the
Efimov constant that determines the geometric scaling of the energy
levels, viz., $\gamma$ in Eq. (\ref{eq:10}).  Although $\gamma$ is not
universal, it is still a characteristic number for the effect, and the
value Efimov determined for fermions is $\gamma=(22.7)^{-2}$. The analogy
with DNA seems to provide a different way of having analogue behaviour, using a 
polymer-based Monte Carlo approach, in particular here we use the flatPERM method introduced by Prellberg and Krawczyk\cite{perm}.

For this purpose, we evaluated the free energy for three chains with
the same interaction $\epsilon_1=\epsilon_2=\epsilon_c^{(2)}$ so that any
pair would be at its critical point.  The free energy has been
evaluated for lengths up to 10,000.

To test the quality of the numerical data obtained from PERM, we check the
nature at $\epsilon=\epsilon_c^{(2)}=1.07726$.  In fact, the simulations done 
 at this point will not give critical point data, as there
is still a small shift $\delta T \neq 0$ due to finite size effects. 

Let us first derive the expected two-chain and three-chain scaling behaviour in order to compare with our flatPERM data. 

For a contact interaction, $V({\bf r}_i(t) -{\bf r}_j(t))= v_0
\delta({\bf r}_i(t) -{\bf r}_j(t)), v_0<0 $ in Eq. (\ref{eq:1}),
standard dimensional analysis tells us
$[s]=[L]^2$, and $[v_0]=[L]^{d-2}$, where $[L]$ denotes the
dimension of length.   For $d>2$, the two chain melting is described by a
Renormalization group fixed point $u^*=2\pi\varepsilon$, where
$\varepsilon=2-d$ and $u$ is the renormalized dimensionless coupling constant
with the bare value $u_0=v_0L^{\varepsilon}$.
In $d=3$, the melting point is the unstable fixed point $u^*=-2\pi \
(\epsilon=-1)$.  
At this fixed point, we  associate the exponent for length $\xi$ as
\begin{equation}
\xi\sim |\Delta T|^{-\phi}, \quad \phi=\frac{1}{|\varepsilon|},
\end{equation}
where $\Delta T=\frac{v_0-v_c}{v_c}$ is the deviation from the melting
point,  so that in $d=3$, 
\begin{equation}
  \label{eq:14}
  \xi_{\parallel} \sim  |\Delta T|^{-2}.
\end{equation}
Here $\xi$ is the length in space while $\xi_{\parallel}$ is along the
chain (number of bases).

The partition function is that of two or three interacting chains,
free at one end but tied together at the other.  This configuration
goes by the name of "survival" partition functions of ``vicious
walkers".  At the unstable fixed point, the behaviour of the finite length
$p$-chain partition function  is of the form
\begin{equation}
  \label{eq:13}
  Z_N^{(p)}\sim \mu^N N^{-\psi_p},
\end{equation}
where  the exponent  $\psi_p$  is
given by \cite{smpre}  
\begin{eqnarray}
\psi_p&=&\eta_p/2, \hspace{1cm}\hfill\\ 
 \eta_p&=&\left\{  \begin{array}{ll}
                         \varepsilon& {\rm for}\  p=2,( {\rm exact})\\ 
                        3\varepsilon+ 3 \ln(3/4)\; \varepsilon^2
                        +O(\varepsilon^3) & {\rm for\ } p=3,\hspace{1cm}\hfill 
                         \end{array}.\right . \label{eq:15}
\end{eqnarray} 
We now use it for $d=3,$ i.e., $\varepsilon=-1$.  At this fixed point,  
we find
$\psi_2=-1/2$, for two chains at the melting point.
For the critical contribution to the three-chain free energy,  
a direct sum gives $\psi_3=-1.9315$ which we approximate as $-2$.

Combining Eqs. (\ref{eq:14}) and (\ref{eq:15}), the scaling for a
small $\delta T$ is (see also Ref \cite{causo})
\begin{equation}
 \label{eq:11}
  Z_N^{(2)}\sim (2d)^{2N} N^{1/2} \ {\cal G}(\delta T\; N^{1/2}),
\end{equation}
For a small fixed $\delta T,\ \ {\cal G}(x)\approx  a'+ b' x.$ 
Therefore, with $\mu=2d$ (for a random walk on a $d$-dimensional
cubic lattice), we have
\begin{equation}
  \label{eq:12}
 F_N^{(2)} \equiv  \ln \left(\frac{Z_N^{(2)}}{\mu^{2N}}\right) \approx a
 +|\psi_2| \ln N + b N^{1/2}.
\end{equation}
For the three chain case, the critical point contribution to the free energy
\begin{equation}
  \label{eq:16}
 F_N^{(3)}\equiv  \ln \left(\frac{Z_N^{(3)}}{\mu^{3N}}\right) \approx a_3
 +|\psi_3| \ln N + b_3 N^{1/2},
\end{equation}
without the Efimov effect.  The Efimov tower (Eq. (\ref{eq:10})) would
lead to a different class of terms of the type
\begin{equation}
  \label{eq:17}
 F_{\rm efi}\sim N |E_0|+\sum c_j e^{-N k_j},
\end{equation}
where $k_j$'s come from the energy gaps of the tower. The absolute
sign for $E_0$ is required because we are actually writing the
expression for $\ln Z$. Combining the two different contributions we
have a form 
\begin{equation}
  \label{eq:18}
 F_N^{(3)}= f_3 N +|\psi_3| \ln N + b_3 N^{1/2}+ a_3 +  \sum c_j \exp(-N k_j).
\end{equation}
Note that, even though $F_N^{(2)}$ has no $O(N)$ term,  a linear
term in $F_N^{(3)}$ with $f_3>0$  is a signature of the Efimov effect.

We use this RG based formula to fit numerically calculated free energy.

\section{Comparison with flatPERM data}
We determined the free energies of the 2-chain and the 3-chain systems
at the two chain melting point $\epsilon_c^{(2)}=1.07726$) for lengths up to
10,000 for $10^8$ iterations

\begin{figure}
\includegraphics[width=\linewidth]{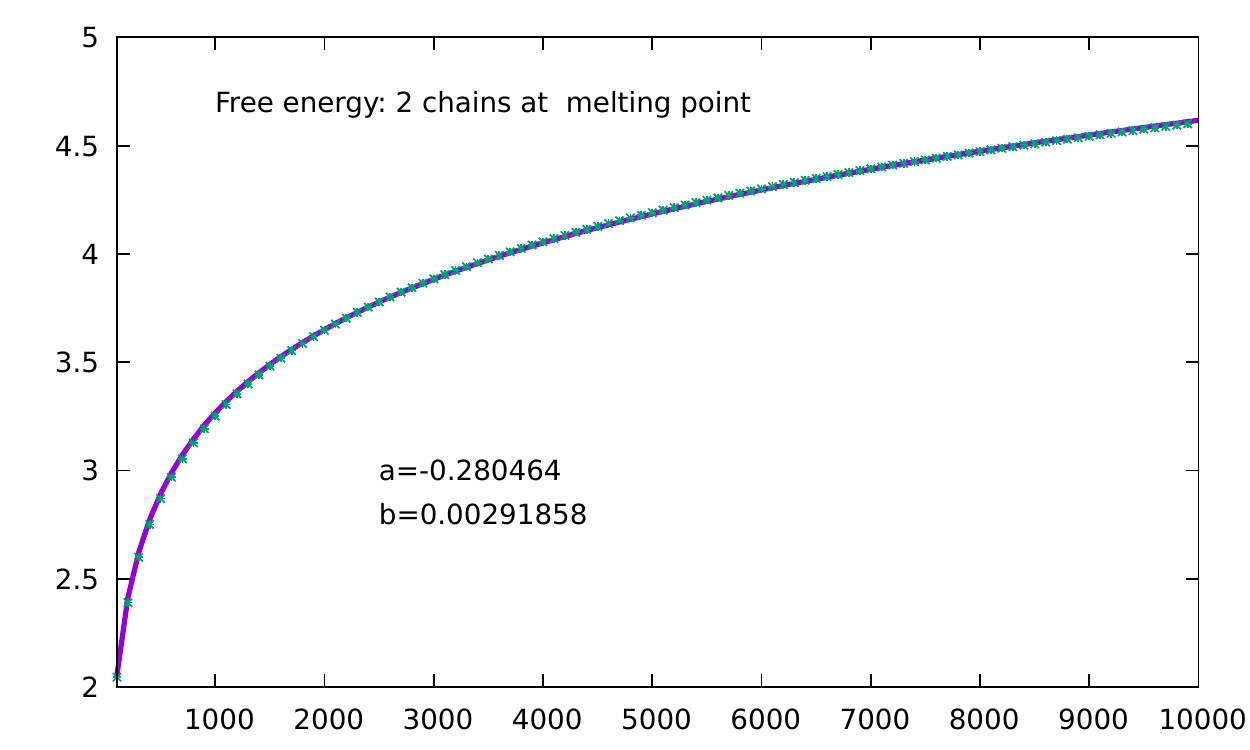}
\caption{Length dependence of the critical point free energy for two
  chains.  RG functional form, Eq. \ref{eq:12}, (solid line) fits
  the data from PERM evaluated at $\epsilon=1.07726$ (blue
  dots), with $a=0.280464, b=0.00292$.}\label{fig:2crit}

\end{figure}

Fig. \ref{fig:2crit} shows a good fit of  Eq. \ref{eq:2} to the
critical point data, with parameters as noted in the caption.  
The good fit is an  assurance of the good quality of data.

Armed with the success for two chains, we try to fit to equation~\ref{eq:18}. The best fit to the data was provided by the following form:
\begin{eqnarray}\label{eq:19}
F_{3}&=& f_3 N + 2 \ln N + b_3 N^{0.5} +a_3,\\\label{eq:20}
\delta F&=&c_0e^{-kN}+c_1e^{-\gamma kN}+c_2e^{-\gamma^2 kN}+c_3e^{-\gamma^3 kN}.
\end{eqnarray}
with $f_3,a_3, b_3$ and the $c_i$ as fitting parameters to the free energy data for
three chains at $\varepsilon_c$.

\begin{figure}
\includegraphics[width=\linewidth]{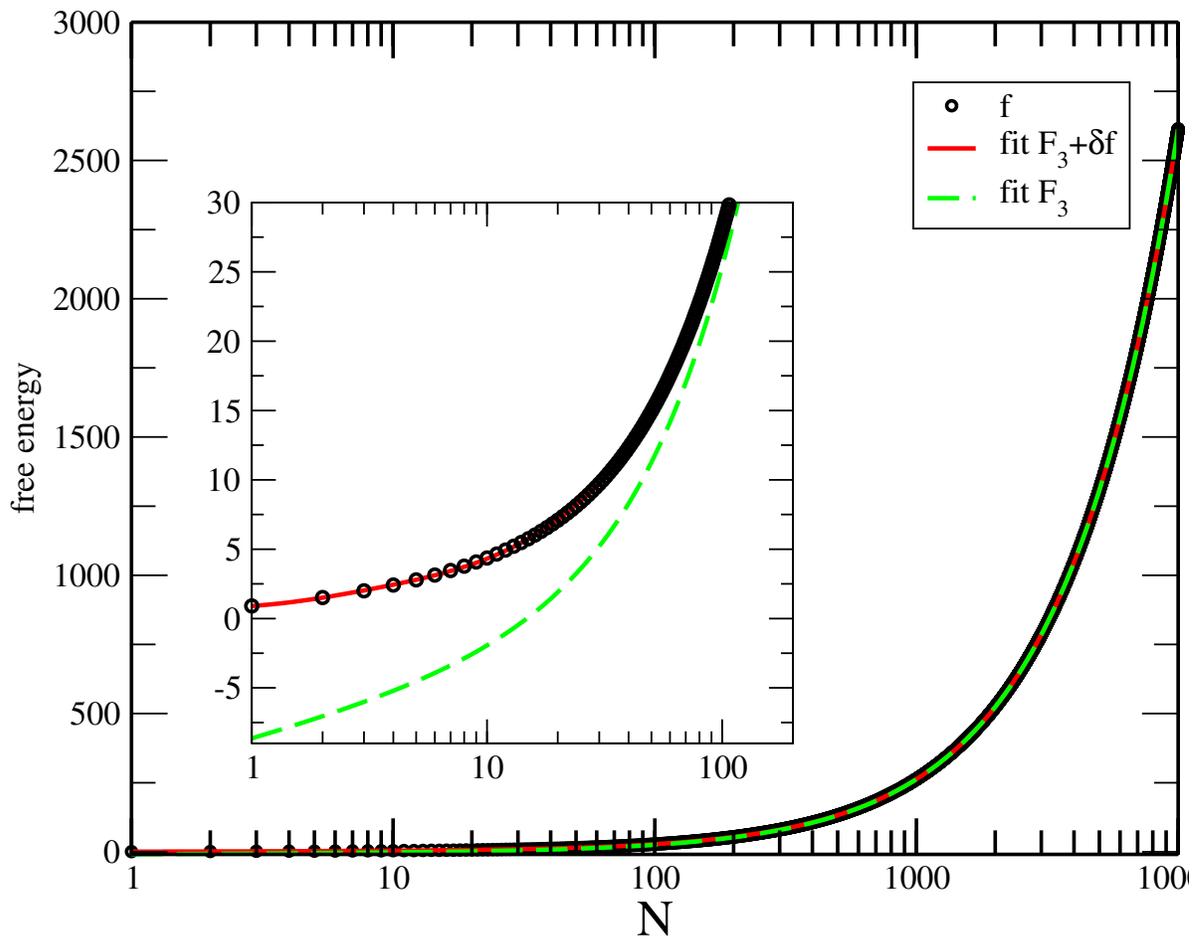}
\caption{Length dependence of 3-chain free energy. The red dots are
  the data points obtained by subtracting Eq. (\ref{eq:19}) from the
  calculated free energy. Different regions are fitted with exponentials.
The green curve fits the small $N$ part,  while the blue curve fits
the intermediate decade.}\label{efi_fit}
\end{figure}

A fit to $F_3$, given by Eq.~\ref{eq:19}, with reduced
$\chi^2=0.053$ gave $f_3=0.26182, b_3=-0.14524,a_3=-7.9429$ with
errors in the last digit, as estimated by the fitting program of
{\small GNUPLOT}.  Fitting to $F_{3}+\delta F$
(Eqs.~\ref{eq:19},~\ref{eq:20}) gives a fit with the parameters
reported in Table~\ref{efi_fit_tab}. The two fits are shown in
Fig.~\ref{efi_fit}. To see the difference in fit, we need to look at
short chain lengths, and the exponential terms are required to ensure
a good fit with a
$\gamma\approx 0.107$.

\begin{table}[h]
\begin{center}
\begin{tabular}{|c|c|c|c|c|}
\hline
$f_3$ & $ b_3$ &  $a_3$ & $c_0$ & $c_1$\\
\hline
 0.261660 &  -0.121678 & -8.781304 & 4.012730 & 2.78526 \\
\hline
 $c_2$&$c_3$ &  $\gamma$ &  $k$ &\\
 \hline
 4.165424 &  2.727346 &0.107447&  2.036299 &\\
 \hline
 
\end{tabular}
\end{center}
\caption{Best fit parameters, which lead to the fit shown in Fig~\ref{efi_fit}.}\label{efi_fit_tab}
\end{table}

This fit shows the existence of at least three bound states, but the energy gaps, determined by $\delta E_i=\gamma^i k$ does not fit the Efimov tower prediction, where we would have expected $\delta E_i=(1-\gamma^i) f_3$, so whilst we confirm that we have multiple bound states in the three-bound phase, where any two chains would not be bound, we do not confirm the Efimov tower. This could be either an indication that the behaviour is different, or rather the finite walk aspect of our investigation does not allow us to see all the possible bound states.

\section{Discussion}

In this article we presented results for the DNA analogue of Efimov-type scaling 
consisting of three-stranded DNA
modelled by three random walkers starting from a common origin, and
interacting only when two walks meet an equal number of steps from the
origin, and the generalised model is studied for its phase diagram in $\varepsilon_1$ and $\varepsilon_2$.

At the equivalent point to the Efimov point ($\varepsilon_1=\varepsilon_2=1.07726$) we find three bound states, and excellent fitting to the free energy from length 0 to 10000, but these states don't follow the Efimov tower structure expected. Whilst the different energy states give the finite-size scaling behaviour of our system, it is not guaranteed  that all will contribute. In the quantum system, we are looking at the eigenvalues of the time evolution operator, which in our case would correspond to a transfer matrix which adds a step to each of the DNA molecules. This can be seen as three interacting partially directed walks in 3+1 dimensions. In a future work we will look at this transfer matrix to see if it is possible to extract the eigenvalue structure directly.


\begin{thebibliography}{99}
\bibitem{efimov1}  V. Efimov, Phys. Lett.B 33,563 (1970).

\bibitem{efimov2}  V. Efimov, Sov. J. Nucl. Phys.12, 589 (1971).

\bibitem{Braaten}  E.  Braaten  and  H.  W.  Hammer,  Phys.  Rep.428,
  259(2006).

\bibitem{kraem}  T. Kraemeret. al., Nature (London) 440, 315 (2006).

\bibitem{zacca} M. Zaccantiet. al., Nat. Phys. 5, 586 (2009).

\bibitem{pires}  R. Pires, J. Ulmanis, S. H Ìafner, M. Repp, A. Arias,
  E.D. Kuhnle, and M. Weidem Ìuller, Phys. Rev. Letts.112, 250404
  (2014).
\bibitem{maji1}  J. Maji, S. M. Bhattacharjee, F. Seno, and
  A. Trovato,New J. Phys.12, 083057 (2010) .

\bibitem{causo}M.  S.  Causo,  B.  Coluzzi,  and  P.  Grassberger, 
Phys.  Rev.  E62,  3958(1999). 

\bibitem{watcri} J. D. Watson and F. H. C. Crick, Nature 171, 737 (1953)


\bibitem{scatt} For scattering, one may take the limit of scattering
  length going to infinity.  The tuning of the potential to the
  critical value of zero energy bound state is done for cold atoms via
  Feschbak resonance.
\bibitem{hammer}E. Brateen and H. W. Hammer,Phys. Rep.428, 259(2006).
\bibitem{pal}   T. Pal, P. Sadhukhan, and S. M. Bhattacharjee   Phys. Rev. Lett. {`bf 110}, 028105 (2013)
\bibitem{perm}  T. Prellberg and J. Krawczyk, Phys. Rev. Lett. {\bf 92}, 120602 

\bibitem{smpre} S. Mukherji and S. M. Bhattacharjee, Phys. Rev. {\bf E 48}, 3427
  (1993).

\bibitem{hier}Jaya Maji and S. M. Bhattacharjee,   Phys. Rev. {\bf E 86},
  041147 (2012) 
  
\bibitem{night1976} M.P Nightingale, Physica A {\bf 83} 561 (1976)
  

\end{thebibliography}
\end{document}